\documentclass[12pt,a4paper]{article}
\usepackage{epsfig}
\begin{document}
\title{Production of the neutral top-pion $\pi_{t}^{0}$ in
association with a high-$p_{T}$ jet at the $LHC$}
\author{Shi-Hai Zhu, Chong-Xing Yue, Wei Liu, and Li Ding  \\
{\small Department of Physics, Liaoning  Normal University, Dalian
116029, P. R. China}
\thanks{E-mail:cxyue@lnnu.edu.cn}}
\date{\today}

\maketitle
\begin{abstract}
In the framework of the topcolor-assisted technicolor $(TC2)$ model,
we study production of the neutral top-pion $\pi_{t}^{0}$ in
association with a high-$p_{T}$ jet at the $LHC$, which proceeds via
the partonic processes $gg\longrightarrow \pi_{t}^{0}g$,
$gq\longrightarrow \pi_{t}^{0}q$, $q\bar{q}\longrightarrow
\pi_{t}^{0}g$, $gb(\bar{b})\longrightarrow \pi_{t}^{0}b(\bar{b})$,
and $b\bar{b}\longrightarrow \pi_{t}^{0}g$. We find that it is very
challenging to detect the neutral top-pion $\pi_{t}^{0}$ via the
process $pp\longrightarrow \pi_{t}^{0}+jet+X\rightarrow
t\bar{t}+jet+X$, while the possible signatures of $\pi_{t}^{0}$
might be detected via the process $pp\longrightarrow
\pi_{t}^{0}+jet+X\rightarrow(\bar{t}c+t\bar{c})+jet+X$ at the $LHC$.

\vspace{1cm}
\end{abstract}

\newpage
\noindent{\bf 1. Introduction}

The Higgs mechanism for the electroweak symmetry breaking $(EWSB)$
is still the untested part of the standard model $(SM)$. Searching
for the $SM$ Higgs boson is one of the main tasks of the forthcoming
Large Hadron Collider $(LHC)$, which has considerable capability to
discover and measure almost all of its quantum properties [1].
However, if the $LHC$ finds evidence for a new scalar state, it may
not necessarily be the $SM$ Higgs boson. Most of new physics models
beyond the $SM$ predict the existence of new scalar states. These
new particles may has production cross sections and branching ratios
which differ from those of the $SM$ Higgs boson. Distinguishing the
various new physics scenarios is an important task for current and
near future high energy collider experiments. Thus, studying the
production and decay of the new scalar states at the $LHC$ is of
special interest.

Due to the large gluon luminosity, the main production mechanism for
a scalar Higgs boson at the $LHC$ is the partonic gluon fusion
process $gg\rightarrow H$ [2], which is the so-called inclusive
single Higgs boson production channel. In order to fully explore the
Higgs detection capabilities of the $LHC$, one should investigate
more exclusive channels, like e.g. Higgs production in association
with a high-$p_{T}$ hadronic jet [3]. The main advantage of this
channel is the richer kinematical structure of the events which
allows for refined cuts increasing the signal-to-background ratio.
So far, this production channel has been extensively studied in the
$SM$ [4,5]. In the minimal supersymmetric standard model $(MSSM)$,
the analogous process, i.e. scalar Higgs production in association
with a high-$p_{T}$ jet was also extensively studied in Refs.[6,7].

Among various kinds of dynamical $EWSB$ theories, the topcolor
scenario is attractive because it can explain the large top quark
mass and provides a possible $EWSB$ mechanism [8]. The
topcolor-assisted technicolor $(TC2)$ model [9] is one of the
phenomenologically viable models, which has all essential features
of the topcolor scenario. This model predicts three $CP$ odd
top-pions ($\pi_{t}^{0}$,$ \pi_{t}^{\pm}$) with large Yukawa
couplings to the third family. The aim of this paper is to consider
the production of the neutral top-pion $\pi_{t}^{0}$ associated with
a high-$p_{T}$ jet and compare our results with those for the Higgs
boson from the $SM$ or the $MSSM$. We hope that our work can help
the upcoming $LHC$ to test topcolor scenario and to differentiate
various kinds of new physics models.

In the rest of this paper, we will give our results in detail. In
section 2, we will calculate the production cross section of the
hadronic process $pp\longrightarrow \pi_{t}^{0}+jet+X$ and give a
simply phenomenological analysis at the $LHC$. Our conclusion is
represented in section 3.

\noindent{\bf 2. Production of the neutral top-pion $\pi_{t}^{0}$
associated with a high-$p_{T}$ jet}

In the $TC2$ model [9], topcolor interactions, which are not
flavor-universal and mainly couple to third generation fermions,
generally generate small contributions to $EWSB$ and give rise to
the main part of the top quark mass. Thus, the top-pions $
\pi_{t}^{0,\pm}$ have large Yukawa couplings to the third generation
fermions. Such features can result in large tree-level flavor
changing couplings of the top-pions to the fermions when one writes
the interactions in the fermion mass eigen-basis. Just as for the
$SM$ Higgs boson, the couplings of the top-pion to a pair of quarks
are proportion to the quark masses. The explicit form for the
couplings of the neutral top-pion $ \pi_{t}^{0}$ to quarks, which
are related to our calculation, can be written as [9,10]:
\begin{eqnarray}
\frac{im_{t}}{\sqrt{2}F_{t}}\frac{\sqrt{\nu_{W}^{2}-F_{t}^{2}}}
{\nu_{W}}[k_{UR}^{tt}k_{UL}^{tt^{*}}\bar{t}\gamma^{5}t
\pi_{t}^{0}+\frac{m_{b}-m_{b}'}{m_{t}}\bar{b}\gamma^{5}b
\pi_{t}^{0}+k_{UR}^{tc^{*}}k_{UL}^{tt}\bar{t}P_{R}c \pi_{t}^{0}],
\end{eqnarray}
where $\nu_{W}=\nu/\sqrt{2}\approx174GeV$, $P_{R}=(1+\gamma^{5})/2$
is the right-handed projection operator, $F_{t}\approx50GeV$ is the
top-pion decay constant, and $m_{b}'\approx0.1\varepsilon m_{t}$ is
the part of the bottom quark mass generated by extended technicolor
interactions. $k_{UL(R)}$ are rotation matrices that diagonalize the
up-quark mass matrix $M_{U}$ for which the Cabibbo-Kobayashi-Maskawa
($CKM$) matrix is defined as $V_{CKM}=k_{UL}^{+}k_{DL}$. To yield a
realistic form of $V_{CKM}$, it has been shown that the values of
the matrix elements $k_{UL(R)}^{ij}$ can be taken as [10]:
\begin{eqnarray}k_{UL}^{tt}\approx1,\hspace*{2cm} k_{UR}^{tt}=1-\varepsilon,
\hspace*{2cm}k_{UR}^{tc}\leq\sqrt{2\varepsilon-\varepsilon^{2}}.
\end{eqnarray}
In our numerical estimation, we will take
$k_{UR}^{tc}=\sqrt{2\varepsilon-\varepsilon^{2}}$ and take
$\varepsilon$ as a free parameter, which is assumed to be in the
range of $0.01\sim0.1$.

\begin{figure}
\begin{center}
\epsfig{file=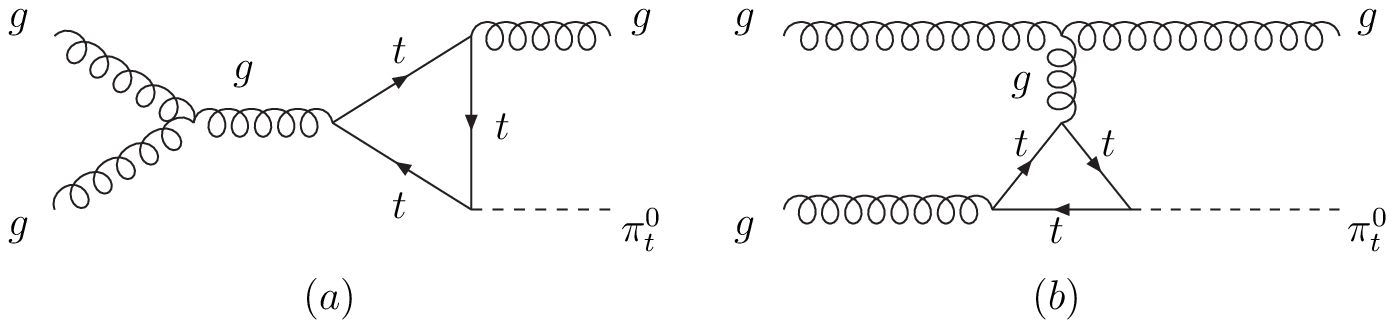,scale=1.1} \epsfig{file=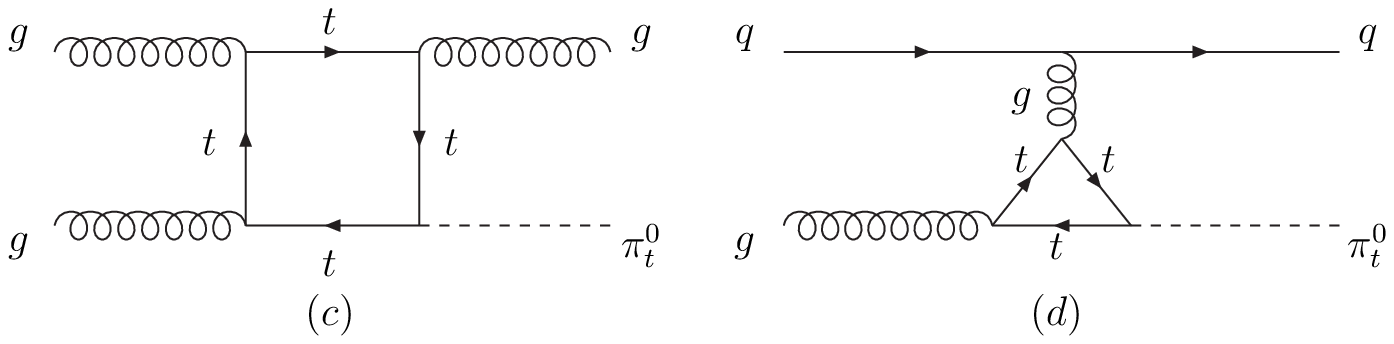,scale=1.1}
\epsfig{file=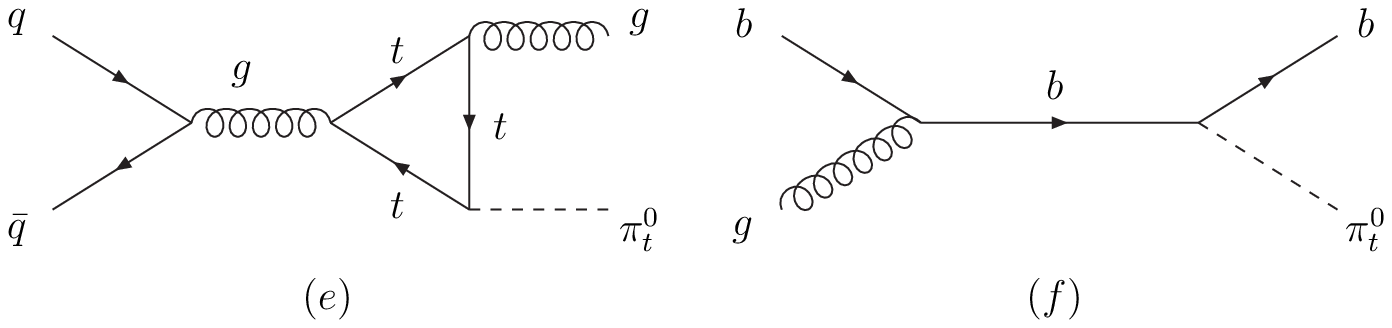,scale=1.1} \epsfig{file=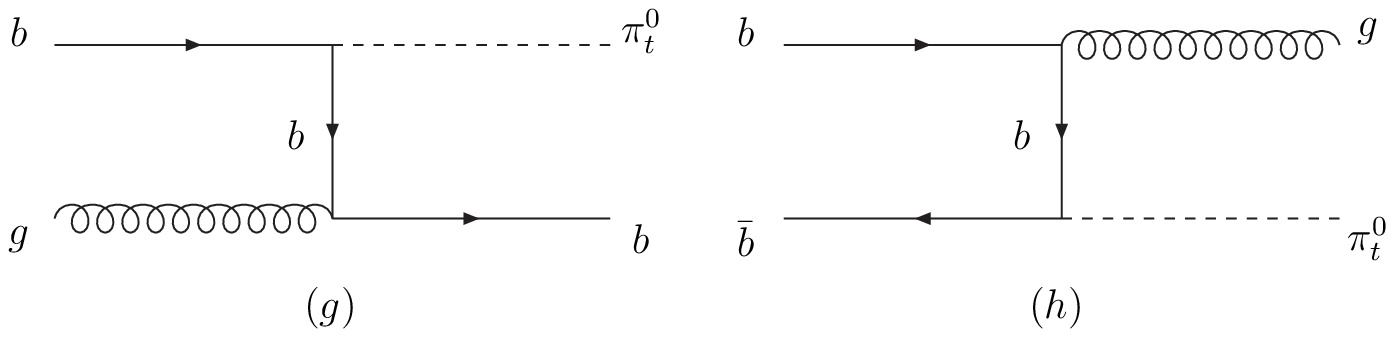,scale=1.1}
\caption{Feynman diagrams for the partonic processes contributing to
the hadronic \hspace*{2cm}process $pp\rightarrow\pi_{t}^{0}+jet+X$
at the leading order. The other diagrams obtained \hspace*{2cm}by
exchanging the gluons or exchanging $\pi_{t}^{0}$ are not shown
here.}
\end{center}
\end{figure}

\newpage
Similar to the Higgs boson predicted by the $SM$ or the $MSSM$, the
neutral top-pion $\pi_{t}^{0}$ can be produced at the $LHC$ in
association with a high-$p_{T}$ jet through three partonic
processes: gluon fusion ($gg\longrightarrow g\pi_{t}^{0}$),
quark-gluon scattering ($q(\bar{q})g\longrightarrow
q(\bar{q})\pi_{t}^{0}$), and quark-antiquark annihilation
($q\bar{q}\longrightarrow g\pi_{t}^{0}$). Although the gluon fusion
and quark-gluon scattering partonic processes give main
contributions to the production cross section for the hadronic
process $pp\longrightarrow \pi_{t}^{0}+jet+X$ at the $LHC$, our
numerical analysis include all of above three processes, which
proceed at one-loop, as shown in $Fig.1 (a)\sim(e)$. Considering the
small value of the decay constant $F_{t}$ and the relatively large
bottom-quark mass, we also consider the contributions of the
tree-level partonic processes $gb\longrightarrow \pi_{t}^{0}b$ and
$b\bar{b}\longrightarrow \pi_{t}^{0}g$, as shown in $Fig.1
(f)\sim(h)$.

The variant amplitudes corresponding to the Feynman diagrams as
shown in $Fig.1$ can be written as:

\begin{eqnarray}
M_{(a)}=&&\frac{i}{4\sqrt{2}\pi^{2}}T^{c_{3}}_{ij}T^{c_{4}}_{ji}f^{c_{1}c_{2}c_{3}}g_{s}^{3}
\frac{m_{t}^{2}(1-\varepsilon)\sqrt{\nu_{W}^{2}-F_{t}^{2}}}
{F_{t}\nu_{W}(p_{1}+p_{2})^{2}}C_{0(a)}
[(p_{2}-p_{1})_{\mu}\epsilon(p_{1})\cdot{}\epsilon(p_{2})\nonumber\\
&&+(-2p_{2}-p_{1})\cdot{}\epsilon(p_{1})
\epsilon_{\mu}(p_{2})+(2p_{1}+p_{2})\cdot{}\epsilon(p_{2})\epsilon_{\mu}(p_{1})]
\epsilon^{\mu\nu\rho\sigma}p_{3\rho}p_{4\sigma}\epsilon_{\nu}(p_{3}),\\
M_{(b)}=&&\frac{i}{4\sqrt{2}\pi^{2}}T^{c_{3}}_{ij}T^{c_{4}}_{ji}f^{c_{1}c_{2}c_{3}}g_{s}^{3}
\frac{m_{t}^{2}(1-\varepsilon)\sqrt{\nu_{W}^{2}-F_{t}^{2}}}
{F_{t}\nu_{W}(p_{1}-p_{3})^{2}}C_{0(b)}
[(-p_{2}-p_{1})\cdot{}\epsilon(p_{3})\epsilon_{\mu}(p_{1})\nonumber\\
&&+(p_{2}-p_{3})\cdot{}\epsilon(p_{1})
\epsilon_{\mu}(p_{3})+(p_{3}+p_{1})_{\mu}\epsilon(p_{1})\cdot{}\epsilon(p_{3})]
\epsilon^{\mu\nu\rho\sigma}p_{4\rho}(p_{1}-p_{3})_{\sigma}\epsilon_{\nu}(p_{2}),\\
M_{(c)}=&&\frac{i}{4\sqrt{2}\pi^{2}}T^{c_{1}}_{ij}T^{c_{2}}_{jk}T^{c_{3}}_{ki}g_{s}^{3}
\frac{m_{t}(1-\varepsilon)\sqrt{\nu_{W}^{2}-F_{t}^{2}}}
{F_{t}\nu_{W}}\nonumber\\&&(-D_{0(c)}m_{t}^{3}\epsilon^{\mu\nu\rho\sigma}p_{1\sigma}
+D_{0(c)}m_{t}^{3}\epsilon^{\mu\nu\rho\sigma}p_{4\sigma}
-D_{0(c)}m_{t}\epsilon^{\rho\sigma\alpha\beta}p_{1\sigma}p_{2\alpha}p_{4\beta}g^{\mu\nu}\nonumber\\&&
+D_{0(c)}m_{t}\epsilon^{\nu\sigma\alpha\beta}p_{1\sigma}p_{2\alpha}p_{4\beta}g^{\mu\rho}
-D_{0(c)}m_{t}\epsilon^{\nu\rho\sigma\alpha}p_{1}^\mu
p_{2\sigma}p_{4\alpha}
+D_{0(c)}m_{t}\epsilon^{\nu\rho\sigma\alpha}p_{1\sigma}p_{4\alpha}p_{2}^\mu\nonumber\\&&
-D_{0(c)}m_{t}\epsilon^{\nu\rho\sigma\alpha}p_{1\sigma}p_{2\alpha}p_{4}^{\mu}
-D_{0(c)}m_{t}\epsilon^{\mu\sigma\alpha\beta}p_{1\sigma}p_{2\alpha}p_{4\beta}g^{\nu\rho}
+2D^\nu_{(c)}m_{t}\epsilon^{\mu\rho\sigma\alpha}p_{1\sigma}p_{4\alpha}\nonumber\\&&
+2D^\nu_{(c)}m_{t}\epsilon^{\mu\rho\sigma\alpha}p_{2\sigma}p_{4\alpha}
+D_{0(c)}m_{t}\epsilon^{\mu\rho\sigma\alpha}p_{2\sigma}p_{4\alpha}p_{1}^\nu
-D_{0(c)}m_{t}\epsilon^{\mu\rho\sigma\alpha}p_{1\sigma}p_{4\alpha}p_{2}^\nu\nonumber\\&&
+D_{0(c)}m_{t}\epsilon^{\mu\rho\sigma\alpha}p_{1\sigma}p_{2\alpha}p_{4}^\nu
-2D^\nu_{(c)}m_{t}\epsilon^{\mu\rho\sigma\alpha}p_{2\sigma}p_{4\alpha}
-D_{0(c)}m_{t}\epsilon^{\mu\nu\sigma\alpha}p_{2\sigma}p_{4\alpha}p_{1}^\rho\nonumber\\&&
-D_{0(c)}m_{t}\epsilon^{\mu\nu\sigma\alpha}p_{1\sigma}p_{4\alpha}p_{2}^\rho
+D_{0(c)}m_{t}\epsilon^{\mu\nu\sigma\alpha}p_{1\sigma}p_{2\alpha}p_{4}^\rho
+D_{\nu(c)}{}D_{(c)}^\nu m_{t}\epsilon^{\mu\nu\rho\sigma}p_{1\sigma}\nonumber\\&&
-D_{\nu(c)}{}D_{(c)}^\nu m_{t}\epsilon^{\mu\nu\rho\sigma}p_{4\sigma}
-2D_{\nu(c)}\cdot{}p_{2}m_{t}\epsilon^{\mu\nu\rho\sigma}p_{1\sigma}
+2D_{\nu(c)}\cdot{}p_{2}m_{t}\epsilon^{\mu\nu\rho\sigma}p_{4\sigma}\nonumber\\&&
-2D_{\nu(c)}\cdot{}p_{4}m_{t}\epsilon^{\mu\nu\rho\sigma}p_{2\sigma}
+D_{0(c)}p_{1}\cdot{}p_{2}m_{t}\epsilon^{\mu\nu\rho\sigma}p_{4\sigma}
-D_{0(c)}p_{1}\cdot{}p_{4}m_{t}\epsilon^{\mu\nu\rho\sigma}p_{2\sigma}\nonumber\\&&
+D_{0(c)}p_{2}\cdot{}p_{2}m_{t}\epsilon^{\mu\nu\rho\sigma}p_{1\sigma}
-D_{0(c)}p_{2}\cdot{}p_{4}m_{t}\epsilon^{\mu\nu\rho\sigma}p_{1\sigma})
\epsilon_{\mu}(p_{1})\epsilon_{\nu}(p_{2})\epsilon_{\rho}(p_{3})\\
M_{(d)}=&&\frac{1}{4\sqrt{2}\pi^{2}}T^{c_{1}}_{ij}T^{c_{1}}_{kl}T^{c_{2}}_{lk}g_{s}^{3}
\frac{m_{t}^{2}(1-\varepsilon)\sqrt{\nu_{W}^{2}-F_{t}^{2}}}
{F_{t}\nu_{W}(p_{1}-p_{3})^{2}}C_{0(d)}
\bar{u}(p_{3})\gamma^{\mu}u(p_{1})g_{\mu\nu}\nonumber\\&&
\epsilon^{\nu\lambda\rho\sigma}p_{4\rho}(p_{1}-p_{3})_{\sigma}
\epsilon_{\lambda}(p_{2}),\\
M_{(e)}=&&\frac{1}{4\sqrt{2}\pi^{2}}T^{c_{1}}_{ij}T^{c_{1}}_{kl}T^{c_{2}}_{lk}g_{s}^{3}
\frac{m_{t}^{2}(1-\varepsilon)\sqrt{\nu_{W}^{2}-F_{t}^{2}}}
{F_{t}\nu_{W}(p_{1}+p_{2})^{2}}C_{0(e)}
\bar{u}(p_{2})\gamma^{\mu}v(p_{1})g_{\mu\nu}\nonumber\\&&
\epsilon^{\nu\lambda\rho\sigma}p_{3\rho}p_{4\sigma}
\epsilon_{\lambda}(p_{3}),\\
M_{(f)}=&&\frac{1}{\sqrt{2}}T^{c_{1}}_{ij}g_{s}\frac{1}{(p_{1}+p_{2})^{2}-m_{b}^{2}}
\frac{m_{b}-m_{b}^{'}}{F_{t}}
\frac{\sqrt{\nu_{W}^{2}-F_{t}^{2}}}{\nu_{W}}\nonumber\\&&
\bar{u}(p_{3})\gamma^{5}(\not{p_{1}}+\not{p_{2}}-m_{b})
\gamma^{\mu}u(p_{1})\epsilon_{\mu}(p_{2}),\\
M_{(g)}=&&\frac{1}{\sqrt{2}}T^{c_{1}}_{ij}g_{s}\frac{1}{(p_{1}-p_{3})^{2}-m_{b}^{2}}
\frac{m_{b}-m_{b}^{'}}{F_{t}}
\frac{\sqrt{\nu_{W}^{2}-F_{t}^{2}}}{\nu_{W}}\nonumber\\&&
\bar{u}(p_{4})\gamma^{\mu}(\not{p_{1}}-\not{p_{3}}-m_{b})
\gamma^{5}u(p_{1})\epsilon_{\mu}(p_{2}),\\
M_{(h)}=&&\frac{1}{\sqrt{2}}T^{c_{1}}_{ij}g_{s}\frac{1}{(p_{1}-p_{3})^{2}-m_{b}^{2}}
\frac{m_{b}-m_{b}^{'}}{F_{t}}
\frac{\sqrt{\nu_{W}^{2}-F_{t}^{2}}}{\nu_{W}}\nonumber\\&&
\bar{v}(p_{2})\gamma^{\mu}(\not{p_{1}}-\not{p_{3}}-m_{b})
\gamma^{5}u(p_{1})\epsilon_{\mu}(p_{4}).
\end{eqnarray}
Here $p_{1},p_{2}$ are the momenta of the incoming states, and
$p_{3},p_{4}$ are the momenta of the outgoing final states. The
$T^{c}_{ij}$ are the $SU(3)$ color matrices and the
$f^{c_{1}c_{2}c_{3}}$ are the antisymmetric $SU(3)$ structure
constants in which $i,j$ are the color indices and
$c_{1},c_{2},c_{3}$ are the indices of gluon. The three-point and
four-point standard functions $C_0,D_0,D_1,D_{\nu}$ [11,12] for
different Feynman diagrams  are defined as:
\begin{eqnarray*}
&&C_{0(a)}=C_{0(a)}(p_{1}+p_{2},-p_{3},m_{t},m_{t},m_{t}),
C_{0(b)}=C_{0(b)}(p_{1}-p_{3},p_{2},m_{t},m_{t},m_{t}),\\
&&C_{0(d)}=C_{0(d)}(p_{1}-p_{3},p_{2},m_{t},m_{t},m_{t}),
C_{0(e)}=C_{0(e)}(p_{1}+p_{2},-p_{3},m_{t},m_{t},m_{t});\\
&&D_{0(c)}=D_{0(c)}(p_{1},p_{2},-p_{3},m_{t},m_{t},m_{t},m_{t}),
D_{1(c)}=D_{1(c)}(p_{1},p_{2},-p_{3},m_{t},m_{t},m_{t},m_{t}),\\
&&D_{\nu(c)}=p_{1\nu}*D_{1(c)}(1)+p_{2\nu}*D_{1(c)}(2)+p_{3\nu}*D_{1(c)}(3).\\
\end{eqnarray*}
Each loop diagram is composed of some scalar loop functions, which
are calculated by using LoopTools [12].

The hadronic cross section at the $LHC$ is obtained by convoluting
the partonic cross sections with the parton distribution functions
$(PDFs)$. In our numerical calculation, we will use
$CTEQ6L\hspace*{0.2cm}PDFs$ [13] for the gluon and quark $PDFs$. The
renormalization scale $\mu_{R}$ and the factorization scale
$\mu_{F}$ are chosen to be $\mu_{R}=\mu_{F}=m_{\pi_{t}}$ for the
gluons and the light quarks, and to be
$\mu_{R}=\mu_{F}=m_{\pi_{t}}/4$ for the bottom-quark, in which
$m_{\pi_{t}}$ is the mass of the neutral top-pion $\pi_{t}^{0}$. To
make our predictions more realistic and high-$p_{T}$ jet not too
close to the beam axis, we require that the transverse momentum
$p_{T}$ and pseudorapidity $\eta$ of the hadronic jet satisfy:
$p_{T}>30GeV$ and $|\eta|<4.5$, which have been used in previous
$MSSM$ studies for the $LHC$ [6,7].

\begin{figure}[htb]
\vspace{-0.5cm}
\begin{center}
\epsfig{file=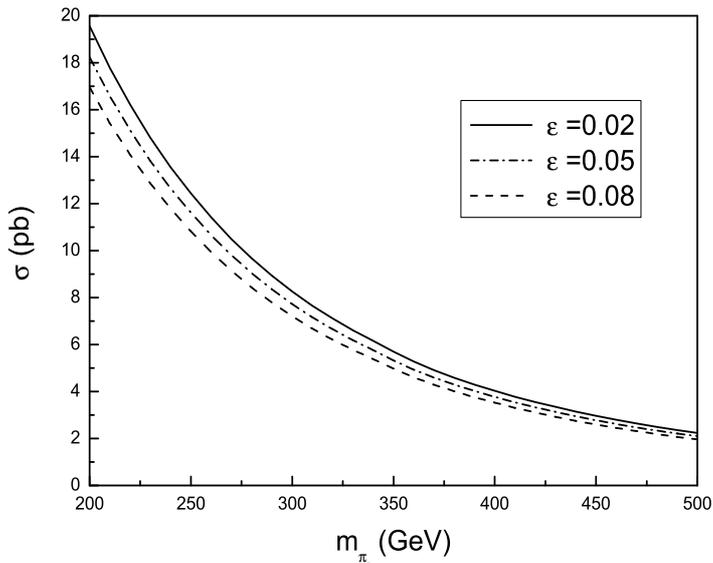,width=300pt,height=250pt} \vspace{-1.0cm}
\hspace{5mm} \caption{The total cross section of the hadronic
process $pp\rightarrow\pi_{t}^{0}+jet+X$ as a function
\hspace*{1.8cm}of the $\pi_{t}^{0}$ mass $m_{\pi_{t}}$ for three
values of the free parameter $\varepsilon$.} \label{ee}
\end{center}
\end{figure}
From the above discussions we can see that the production cross
section $\sigma$ for the hadronic process
$pp\rightarrow\pi_{t}^{0}+jet+X$ is dependent on the free parameters
$\varepsilon$ and $m_{\pi_{t}}$. Similar with Ref.[14], we will
assume that the free parameters $\varepsilon$ and $m_{\pi_{t}}$ are
in the range of $0.01\sim0.1$ and $200GeV\sim500GeV$, respectively.

Our numerical results are shown in $Fig.2$, in which we plot the
cross section $\sigma$ as a function of the mass parameter
$m_{\pi_{t}}$ for three values of the parameter $\varepsilon$. One
can see from $Fig.2$ that $\sigma$ is insensitive to the free
parameter $\varepsilon$. For $\varepsilon=0.05$ and $200GeV\leq
m_{\pi_{t}}\leq 500GeV$, the value of the production cross section
$\sigma$ is in the range of $18.3pb\sim2.1pb$. Observably, if we
assume that the $\pi_{t}^{0}$ mass $m_{\pi_{t}}$ is equal to that of
the $SM$ Higgs boson $H$ or the $MSSM$ Higgs boson $H^{0}$, the
cross section for the production of the neutral top-pion
$\pi_{t}^{0}$ associated with a high-$p_{T}$ is significantly larger
than that of the $SM$ Higgs boson $H$ [4,5] or the $MSSM$ Higgs
boson $H^{0}$ [6,7]. This is because the $\pi_{t}^{0} t
\overline{t}$ coupling is larger than that for the $SM$ Higgs boson
$H$ or the $MSSM$ Higgs boson $H^{0}$.

\begin{figure}[htb]
\vspace*{-0.5cm}
\begin{center}
\epsfig{file=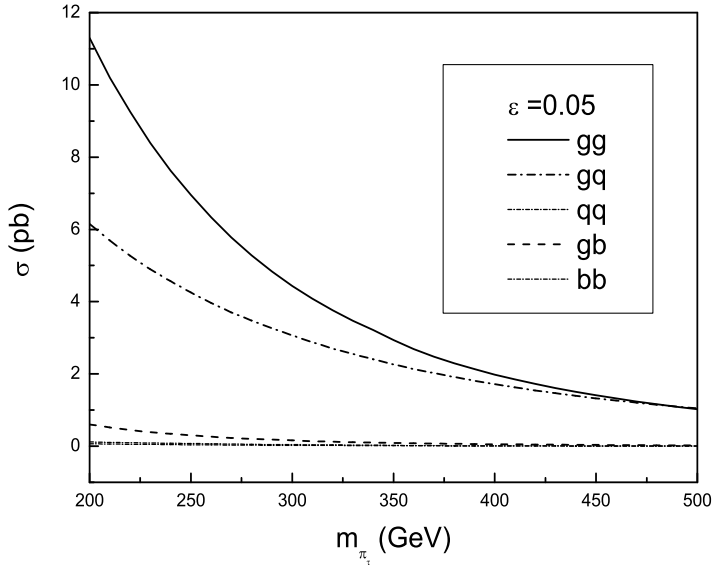,width=300pt,height=250pt} \vspace{-1.0cm}
\hspace{5mm} \caption{The hadronic cross sections for different
partonic processes as function of \hspace*{2.2cm}the $\pi_{t}^{0}$
mass $m_{\pi_{t}}$ for the free parameter
$\varepsilon=0.5$.}\label{ee}
\end{center}
\end{figure}

To see contributions of the different partonic processes to the
total hadronic cross section, we plot the hadronic cross sections of
the partonic processes $gg\rightarrow\pi_{t}^{0}g, qg\rightarrow
q\pi_{t}^{0} (q=u,c,d,s,\bar{u},\bar{c},\bar{d},\bar{s}),
q\bar{q}\rightarrow\pi_{t}^{0}g (q=u,c,d,s),
gb(\bar{b})\longrightarrow \pi_{t}^{0}b(\bar{b})$, and
$b\bar{b}\longrightarrow \pi_{t}^{0}g$ for $\varepsilon$=0.05 in
$Fig.3$. We see that the production of the neutral top-pion
$\pi_{t}^{0}$ in association with a high-$p_{T}$ jet is dominated by
the partonic process $gg\rightarrow\pi_{t}^{0}g$, which is similar
with the Higgs boson production associated with a high-$p_{T}$ in
the $SM$ and the $MSSM$. However, for the $MSSM$ model, the
contributions of the $b\bar{b}$ channel can be significantly large,
depending the free parameters. However, this is not the case for the
$TC2$ model. For $0.02\leq\varepsilon\leq 0.08$ and $200GeV\leq
m_{\pi_{t}}\leq 500GeV$, the hadronic cross section for the partonic
process $b\bar{b}\longrightarrow \pi_{t}^{0}g$ is only in the range
of $1.6fb\sim46fb$, which is several orders of magnitude smaller
than that for the partonic process $gg\rightarrow\pi_{t}^{0}g$.

\begin{figure}[htb]
\vspace*{0.5cm}
\begin{center}
\epsfig{file=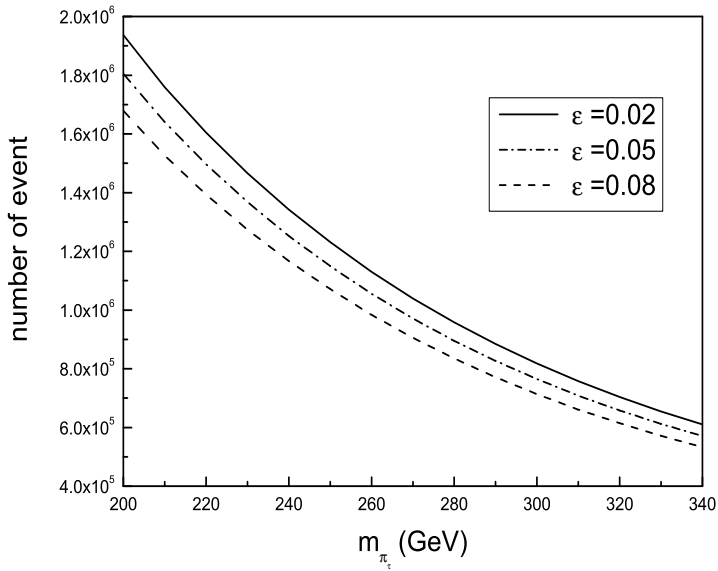,width=300pt,height=250pt} \vspace{-1.0cm}
\hspace{5mm} \caption{The number of the $tc+jet$ event as a function
of the $\pi_{t}^{0}$ mass $m_{\pi_{t}}$ for three \hspace*{1.8cm}
values of the parameter $\varepsilon$.}\label{ee}
\end{center}
\end{figure}
It is well known that the mass of the $SM$ Higgs boson $H$ is
generally smaller than $200GeV$, one can use the decay channels
$H\rightarrow\gamma\gamma$, $H\rightarrow\tau^{+}\tau^{-}$ or
$H\rightarrow W^{+}W^{-}$ to consider  the $SM$ Higgs boson
signatures generated by the hadronic process $pp\longrightarrow
H+jet+X$ at the $LHC$ [5]. For the neutral top-pion $\pi_{t}^{0}$,
its main decay modes are $t\bar{t}, \bar{t}c(t\bar{c}), b\bar{b},
gg$, and $ \gamma\gamma$. For $m_{t}\leq m_{\pi_{t}}\leq 2m_{t}$,
$\pi_{t}^{0}$ mainly decays to $\bar{t}c$ and $t\bar{c}$. It has
been shown that the value of the branching ratio
$Br(\pi_{t}^{0}\rightarrow\bar{t}c+t\bar{c})$ is larger than 90\%
for $m_{\pi_{t}}=250GeV$ and $\varepsilon\geq0.02$ [15]. Thus, for
$m_{t}<m_{\pi_{t}}\leq 2m_{t}$, the production of neutral top-pion
$\pi_{t}^{0}$ associated with a high-$p_{T}$ hadronic jet can easily
transfer to the $tc+jet$ event. This final state generates
characteristic signatures at the $LHC$ experiments. So we further
calculate its production rate. We find that, for
$\varepsilon\leq0.08$ and $m_{\pi_{t}}\leq 350GeV$, the production
cross section of the hadronic process $pp\rightarrow(\bar{t}c+
t\bar{c})+jet+X$ is larger than $19.4pb$. If we assume the yearly
integrated luminosity $\pounds_{int}=100fb^{-1}$ for the $LHC$ with
$\sqrt{s}=14TeV$, then there will be $1.94\times 10^{6}\sim
5.3\times 10^{5} \hspace*{0.2cm} tc+jet$ events to be generated per
year for $0.02\leq\varepsilon\leq0.08$ and $200GeV\leq
m_{\pi_{t}}\leq 340GeV$, as shown in Fig.4.

For the $tc+jet$ event, the peak of the invariant mass distribution
of $tc$ is narrow. To identify $tc$, one needs reconstruct top quark
from its mainly decay mode $Wb$ and the b-tagging and c-tagging are
also needed. Furthermore, in the case of the $W$ hadronic decay, the
$tc+jet$ event will generate the $bjjcj$ final state, while for the
$W$ leptonic decay, it will generate the $bl\nu cj$ final state. For
the former final state, the $SM$ background is $jjjjj$ and the $SM$
backgrounds of the later final state mainly come from the
$t\bar{t}$, $tW$ and $Wjjj$ production process, which have been
analyzed in Ref.[16]. They have shown that suitable kinematical cuts
on the observed particles is more than enough to obtain a clear and
statistically meaningful flavor-changing signal. Thus we expect that
the possible signatures of the neutral top-pion $\pi_{t}^{0}$ might
be detected via the decay channel
$\pi_{t}^{0}\rightarrow\bar{t}c+t\bar{c}$ at the $LHC$ experiments.

For $m_{\pi_{t}}>2m_{t}$, the neutral top-pion $\pi_{t}^{0}$ mainly
decays to $t\bar{t}$ and the hadronic process
$pp\rightarrow\pi_{t}^{0}+jet+X$ can give rise to the $t\bar{t}+jet$
event. Its production rate can reach 15pb for $m_{\pi_{t}}\geq
400GeV$ and $\varepsilon\leq 0.08$. This kind of events have been
calculated at $NLO$ in the $SM$ [17]. It has shown that, for the
renormalization and factorization scales having
$\mu_{R}=\mu_{F}=\mu=m_{t}$, the $NLO$ cross section for
$t\bar{t}+jet$ production at the $LHC$ is larger than $500pb$. Thus,
the production cross section of the $t\bar{t}+jet$ final state
coming from $TC2$ is smaller than that coming from the $SM$ by at
least two orders of magnitude. It is very challenging to detect the
possible signals of $\pi_{t}^{0}$ via the process $pp\longrightarrow
\pi_{t}^{0}+jet+X\rightarrow t\bar{t}+jet+X$.

\noindent{\bf 3. Conclusion}

The production of a scalar state (the $SM$ Higgs boson, the $MSSM$
Higgs boson, etc) associated with a high-$p_{T}$ jet allows for
refined cuts increasing the signal-to-background ratio, which is
considered advantageous for scalar detection even though its
production rate is lower than that for totally inclusive single
scalar state production. In the context of the $TC2$ model, we
consider the production of the neutral top-pion $\pi_{t}^{0}$
accompanied by a high-$p_{T}$ jet at the $LHC$. This production
channel proceeds by the partonic processes $gg\longrightarrow
\pi_{t}^{0}g$, $gq\longrightarrow \pi_{t}^{0}q$,
$q\bar{q}\longrightarrow \pi_{t}^{0}g$, $gb(\bar{b})\longrightarrow
\pi_{t}^{0}b(\bar{b})$, and $b\bar{b}\longrightarrow \pi_{t}^{0}g$.
We find that, for $m_{\pi_{t}}$ equaling to the mass of the scalar
sate predicted by the $MSSM$, the hadronic production cross section
of the process $pp\longrightarrow \pi_{t}^{0}+jet+X$ is much larger
than that for the $MSSM$ scalar state. For
$m_{t}<m_{\pi_{t}}\leq2m_{t}$, the main decay channel is
$\pi_{t}^{0}\rightarrow\bar{t}c+t\bar{c}$. There will be a large
number of the $ tc+jet$ events to be generated which can generate
characteristic signal at the $LHC$ experiment. So we might detect
the possible signatures of the neutral top-pion $\pi_{t}^{0}$ via
the process $pp\longrightarrow
\pi_{t}^{0}+jet+X\rightarrow(\bar{t}c+t\bar{c})+jet+X$ at the $LHC$.

\vspace{1.0cm}

\noindent{\bf Acknowledgments}

Shi-Hai Zhu would like to thank Lei Wang for useful discussions.
This work was supported in part by the National Natural Science
Foundation of China under Grants No.10675057 and Foundation of
Liaoning  Educational Committee(2007T086).

\end{document}